# Moiré-induced bandgap tuning by varying electric dipole in InSe/CuSe vertical heterostructure

Bo Li[1#], Meysam Bagheri Tagani[2#], Sahar Izadi Vishkayi[3], Yumu Yang[1], Jing Wang[1], Qiwei Tian[1], Chen Zhang[1], Li Zhang[1], Long-Jing Yin[1], Yuan Tian[1], Lijie Zhang[1*], and Zhihui Qin[1*]

1. *Key Laboratory for Micro/Nano Optoelectronic Devices of Ministry of Education & Hunan Provincial Key Laboratory of Low-Dimensional Structural Physics and Devices, School of Physics and Electronics, Hunan University, Changsha 410082, China*

2. *Department of Physics, University of Guilan, P.O. Box 41335-1914, Rasht, Iran*

3. *School of Physics, Institute for Research in Fundamental Sciences (IPM), P. O. Box 19395-5531, Tehran, Iran*

B.L. and M.B.T. contributed equally to this work.

*Authors to whom correspondence should be addressed: lijiezhang@hnu.edu.cn;

zhqin@hnu.edu.cn

**Abstract:** The stacked two layered materials with a lattice constant mismatch and/or with twist angle relative to each other can create a moiré pattern, modulating the electronic properties of the pristine materials. Here, we combine scanning tunneling microscopy/spectroscopy and density functional theory calculations to investigate the moiré potential induced bandgap tuning in InSe/CuSe vertical heterostructure synthesized by a two-step of molecular beam epitaxy. Scanning tunneling microscopy measurements demonstrate the heterostructure with a superlattice periodicity of ~3.48 nm and a twist angle of about 11° between the monolayers. Scanning tunneling spectroscopy record on the different stacking sites of the heterostructure reveals the bandgap of the InSe is location-dependent and a variation of 400 meV is observed. Density functional theory calculations reveal that the moiré-induce electric dipole in the monolayer InSe is the key factor for tuning the bandgap. Besides, charge transfer between CuSe and InSe also contributes to the bandgap variation due to its stacking related. We also show that the moiré potential not only can tune the bandgap of InSe but also can vanish the Dirac nodal line of CuSe in some stackings. Our explorations provide valuable information in understanding the electronic properties of the two-dimensional moiré materials.

The emerged moiré materials provide a platform for exploring strong electronic correlations, non-trivial band topology, superconductivity and so on[1,2]. For instance, twist bilayer graphene with magic angle ignites the correlated flat bands induced unconventional superconductor and Mott insulator states[3,4]. Recently, twist bilayer transition metal dichalcogenide (TMDs) are predicted to occur in long-range moiré wavelength[5] and experimentally confirmed in twist homobilayer $WSe_2$[6]. Moiré excitons are reported in twist TMDs systems[7,8]. The electronic properties of twist bilayer semiconducting van der Waals materials are mediated by moiré superlattice. Shabani *et al.* found unexpected large moiré potential in twist heterobilayer $WSe_2$/$MoSe_2$ determined by moiré structure and internal strain[9]. Besides the twist layered semiconductors, bilayer layered metal/metal, semiconductor/metal formed moiré superlattice are also impressive. For instance, Zhao *et al.* investigate the moiré enhanced charge density wave state in metallic 1T-$TiTe_2$/$TiSe_2$[10]. Wu *et al.* investigated the carrier doping modulation from p-type to n-type for $WSe_2$ on Au substrate with van der Waals twisting[11].

InSe is a typical layered post-transition metal chalcogen with high carrier mobility, host quantum Hall effect, and broadband optical response[12-14]. Fu *et al.* found signatures of strong interlayer coupling in InSe, which might smears the moiré potential modulation in twist InSe homostructure[15]. Recently, a monolayer CuSe has been fabricated on Cu(111) substrate by molecular beam epitaxy[16]. The analysis of synthesized monolayer revealed the monolayer CuSe has a honeycomb lattice and hosts Dirac nodal line (DNL) fermions.

Here, motivated by good optical property of InSe and the DNL of CuSe, we investigate the electronic properties of the moiré superlattice of InSe/CuSe heterostructure on Cu(111) substrate. Combined scanning tunneling microscopy/spectroscopy (STM/STS) and density functional theory (DFT) calculations, we find the position dependent bandgap formation of the heterostructure with a twist angle of 11°. DFT calculations reveal that the moiré-induce electrical dipole in the monolayer InSe is the key factor for the bandgap variation.

Our experiments were carried out in a home-built ultrahigh vacuum molecular beam epitaxy (UHV-MBE) chamber equipped with an STM (Unisoku). The base pressure of the system is better than $1\times10^{-10}$ Torr. The Cu(111) substrate was cleaned by several circles of sputtering and annealing. Subsequently, molecular beam In (99.999%, Alfa Aesar) and Se atoms (99.999%, Alfa Aesar) were thermally evaporated from a home-built evaporator onto Cu(111) substrate held at room temperature. All the STM/STS measurements were conducted at room temperature. The STS (*dI/dV-V* curve) measurements were acquired by using a standard lock-in technique (793 Hz, 40-50 mV a.c. bias modulation). The system were carefully calibrate by Si(111)-(7×7) and Au(111) surface.

The Vienna Ab-initio Simulation Package[17] was used to calculate the electronic properties of the heterostructure using density functional theory. The generalized gradient approximation (GGA) was employed with the Perdew-Burke Ernzerhof (PBE) exchange-correlation functional[18]. The plane wave cutoff energy was 450 eV, and the energy self-consistent criteria was $10^{-6}$ eV. To optimize the structures, the atomic positions and lattice sizes are fully relaxed using the conjugate gradient algorithm until the Hellman-Feynman forces drop below $10^{-2}$ eV/ Å. The Brillouin zone was sampled by a 9×9×1 Monkhorst-Pack K-point grid[19], and the DFT-D3 dispersion correction by Grimme is adopted to account for the van der Waals interactions[20]. A vacuum of 20 Å was considered to avoid the interaction between the system with its image.

We deposited ~0.8 monolayer (ML) amount of In atoms onto Cu(111) substrate and annealed the sample mildly to form a few In islands as shown in Fig. 1(a). There are uncovered areas on the substrate marked as Cu in Fig. 1(a). Figure 1(b) shows an atomic resolution STM image of the zoom-in In island highlight by a red square in Fig. 1(a). Subsequently, sub-monolayer of Se atoms were deposited on the substrate and annealed at 400 K, forming multiple InSe hexagonal islands as shown in Fig. 1(c). Meanwhile, the Cu(111) substrate is selenizated as well as the area beneath the In islands by Se intercalation, which is similar with the Pt-Se system[21,22]. However, the surface of the InSe island looks quite rough and we increase the annealing temperature to ~440 K, finding the enlarger InSe islands on the surface (see Fig. 1(d)). The nanopore patterned CuSe is visible as well as some indium clusters. A zoom-in atomic resolutions STM image of CuSe is shown in Fig. 1(e), where triangle and parallelogram shaped nanopores are visible and consistent with the previous work[23,24]. We continue increase the annealing temperature to 473 K, the surface looks quite clean and most of the In clusters are disappeared and nanopore pattered CuSe as well as hexagonal shaped InSe islands are stayed on the surface as shown in Fig. 1(f). On top of the InSe island a moiré pattern is observed, suggesting that the formation of InSe/CuSe heterostructure which will be discussed in detail below.

To gain more information of the surface, we investigated the structural and electronic properties of the obtained InSe. The proposed top and side view of the atomic configuration of monolayer InSe is shown in Fig. 2(a). A typical STM topographic image of large-scale InSe island in Fig. 2(b) reveals that the island possesses an apparent height of 210 pm, providing evidence that the formation of InSe/CuSe vertical heterostructure. Figure 2(c) shows a large-scale STM image of InSe/CuSe moiré superstructure. In order to clarify the randomly shown white dots in Fig. 2(c), we selected different scales of STM images to depict in Fig. S1 in Supplementary Material. Those white dots are atomically resolved in Fig. S1, showing intensity nonuniform as the moiré pattern. The nanopore patterned CuSe below the heterostructure causes these intensity nonuniform. Another possibility is due to the interaction between top layer InSe and impurities/adsorption of dissociative Se or In adatoms, which is similar to that of exhibiting adatoms states in graphene/SiC systems[25]. The zoom-in STM image of the red area in Fig. 2(c) shows that the moiré pattern with a wavelength of 3.48 nm along two different directions (see the line profile in Fig. 2(g)), resulting of the superimposition of CuSe and InSe lattice. The corresponding fast Fourier transform (FFT) in Fig. 2(d) reveals that the outside six moiré spots contributed from the top InSe (purple circles), whereas the inside six moiré spots contributed from the moiré pattern (green circles). Figure 2(e) shows the continue zoom-in of Fig. 2(c), depicting atomic resolution STM image of InSe/CuSe vertical heterostructure with twist angle. Both the moiré superlattice of the heterostructure and the atomic InSe lattice are well resolved (see Fig. 2(e)). The measured lattice periodicity of 0.401 ± 0.001 nm, as shown in line profile in Fig. 2(f), is consistent with the lattice constant of InSe as reported in pervious experimental and theoretical work[15,26]. The lattice constant of CuSe obtained from the inset of Fig 1(e) (~0.41 nm), in consistent with the previous work[24], is considered to calculate the relative twist angle of ~11° with the top InSe layer[27-29]. The estimated twist angle fits nicely with our experimental observations, and the simulated hexagonal moiré pattern is illustrated in Fig. 2(h). It is worth noting that the interplay between the van der Waals interaction energy and the elastic energy at the interface causes atomic reconstruction, which normally only occurs at small twist angles in twist graphene or TMDs systems[30-32]. Here the twist angle is quite large (11º), hampering the formation of reconstruction.

To gain more information of electronic properties of InSe, we carry out the STS measurement on the twist heterostructure, the formed moiré superlattice. Figure 3(a) shows a large-scale STM image of the InSe/CuSe heterostructure. We performed differential conductivity (*dI/dV*) spectra along the line from InSe to CuSe, crossing the interface. Typical *dI/dV* spectra shown in Fig. 3(b) reveals the semiconducting properties of InSe whereas CuSe is metallic[24]. In the following, we focus on the electronic properties of the moiré superlattice.

By recording the *dI/dV* spectra on different sites of the structure as marked by variable color dots in Fig. 3(c). We selected three different positions of the moiré center, bridge and top sites labeled as A, B, and C, respectively. Fig. 3(d)-(f) show the corresponding *dI/dV* spectra on these positions of the moiré superlattice shown with red, blue and cyan curves, respectively. By carefully attract the valence band maximum and conducted band minimum from the correspond logarithms for the *dI/dV-V* spectra as shown in upper panels of Fig. 3(d)-(f). It is easy to find that the bandgaps are modulate with the positions. The selected three typical different positions of the surface showing the bandgap of 1.15, 1.0 and 0.75 eV, respectively. The difference of the observed bandgaps is 400 meV.

To explain the experimental observation, we conduct the DFT calculations to examine the properties of InSe/CuSe heterobilayer. According to initial experimental reports, a heterostructure composed of InSe/CuSe was detected by STM with a twist angle of 11° moiré pattern. With respect to lattice parameters of each monolayer, a moiré heterostructure with lattice constant of 6.27 nm was created as shown in Fig. 4. To unveil the electronic properties of the heterostructure, we investigate the microscopic patterns of the moiré. As it is depicted in the Fig. 4, different stacking of heterobilayer can be detected, that some of them have three-fold rotational symmetry. To examine stacking-dependent electronic properties of heterobilayer, a $\sqrt{3}\times\sqrt{3}$ supercell of monolayers CuSe and InSe is used. The heterostructure has 18 atoms and the strain between the layers is less than 1%. We slide the monolayer InSe along [100] direction to construct different stacking patterns. The value of sliding is denoted by $\Delta$ and lateral shift is done with respect to AB stacking. All considered configurations are fully relaxed with stacking-constraint relaxation method. Some considered stacks with high symmetry are shown in Fig. 4(a). Variation of bandgap of monolayer InSe as a function of stacking is illustrated in blue curve of Fig. 4(c). Our results show that the bandgap is significantly dependent on the arrangement of InSe relative to the substrate. The DFT results confirm the STS observations, demonstrating spatial variations of bandgap. The largest bandgap is observed in $\Delta=0.62$, while the AB stacking has the smallest bandgap. It is of difficult to precisely decide what are the $\Delta$ values for the experimental selected stacking sites. However, by comparing the DFT results with experiment, it can be concluded that the STS results of Fig. 3(d)-(f) is contained in the calculated range. The spatial-dependent bandgap is a direct consequence of change of electrostatic potential by stacking. As it was discussed in Supplementary Material, the InSe can be considered as a bilayer of InSe hexagons coupled by In-In pillars. There is no any potential difference between the layers in free-standing case. By growing InSe on the monolayer CuSe, the symmetry of the structure is disturbed and a potential difference along z-direction is created in the InSe. The potential difference creates an electrical dipole in the structure and modulate its electronic properties. Electrostatic potential of AB stacking is shown in Fig. 4(b). The potential difference between InSe layers is shown by $\Delta V_{InSe}$. Figure 4(c) also illustrates the variation of $\Delta V_{InSe}$ as a function of $\Delta$ (shown in red curve). It is found that the changes of bandgap are in sharp contrast with $\Delta V_{InSe}$. When the $\Delta V_{InSe}$ increases, the bandgap of the monolayer InSe is reduced. Variation of bandgap of monolayer InSe as a function of stacking is illustrated in Fig. 4(c) (blue curve). Our results show that the bandgap is significantly dependent on the arrangement of InSe relative to the substrate. Indeed, the induced local potential difference is a key factor for spatial variation of bandgap that is in consistent with the STS results.

In addition to local potential difference, the charge transfer between the InSe and CuSe is also a key parameter in the bandgap. The charge transfer and the spatial variation of valence and conduction bands in the $\Gamma$ point are depicted in Fig. 5(a). In some atomic configurations, the charge is transferred from CuSe toward InSe, and in some $\Delta$, the direction of charge transfer is reversed. It is worth noting that the charge transfer was calculated for each local stacking that is independent of each other. The charge transfer from the substrate to monolayer InSe was calculated by Bader analysis. In AA stacking that there is a depletion of electron, the valence band of InSe is close to the Fermi level. Indeed, the local moiré potential not only modulates the band gap but also a spatial variation of work function is observed. The positive (negative) value in Fig. 5(a) shows the accumulation (depletion) of charge on InSe. Projected band structure of InSe and CuSe for some registries is drawn in Fig. 5(b) and (c) for different stackings. As it was discussed above the bandgap is dependent on the local moiré pattern. The interlayer interaction between CuSe and InSe not only changes the electronic properties of monolayer InSe, but also affects the ones of monolayer CuSe. As it is illustrated in Fig. 5(b) and 5(c) the type of bandgap, direct or indirect, is dependent on the stacking. So, it is expected that the photoluminance results are dependent on the position. As it is revealed in Fig. S2 the monolayer CuSe hosts a DNL consistent with Ref[16]. Interestingly, the interlayer interaction also affects the electronic properties of monolayer CuSe. As shown in Fig. 5(b), the DNL is vanished in the AA stacking, while it can be detected in AB' stacking.

In summary, we have synthesized InSe/CuSe vertical heterostructure by sequential deposition of In and Se on Cu(111). STM and STS shows the structural and electronic properties of the heterostructure, confirming moiré superlattice with a wavelength of 3.48 nm and a twist 11° with the underlying substrate. STS measurements demonstrate the position-dependent bandgap of the moiré superlattice. DFT calculations reveals the moiré potential caused the varies of the bandgap. Besides, the charge transfer between InSe and CuSe is another factor tuning bandgap.

See the supplementary material for extra experimental data and further details of calculations presented in this work.

This work was supported by the National Natural Science Foundation of China (Grant Nos. 11904094, 51972106, 12174096, and 12174095), the Strategic Priority Research Program of Chinese Academy of Sciences (Grant No. XDB30000000), and the Natural Science Foundation of Hunan Province, China (Grant No: 2021JJ20026). The authors acknowledge the financial support from the Fundamental Research Funds for the Central Universities of China.

AUTHOR DECLARATIONS

Conflict of Interest

The authors have no conflicts to disclose.

DATA AVAILABILITY

The data that support the findings of this study are available from the corresponding author upon reasonable request.

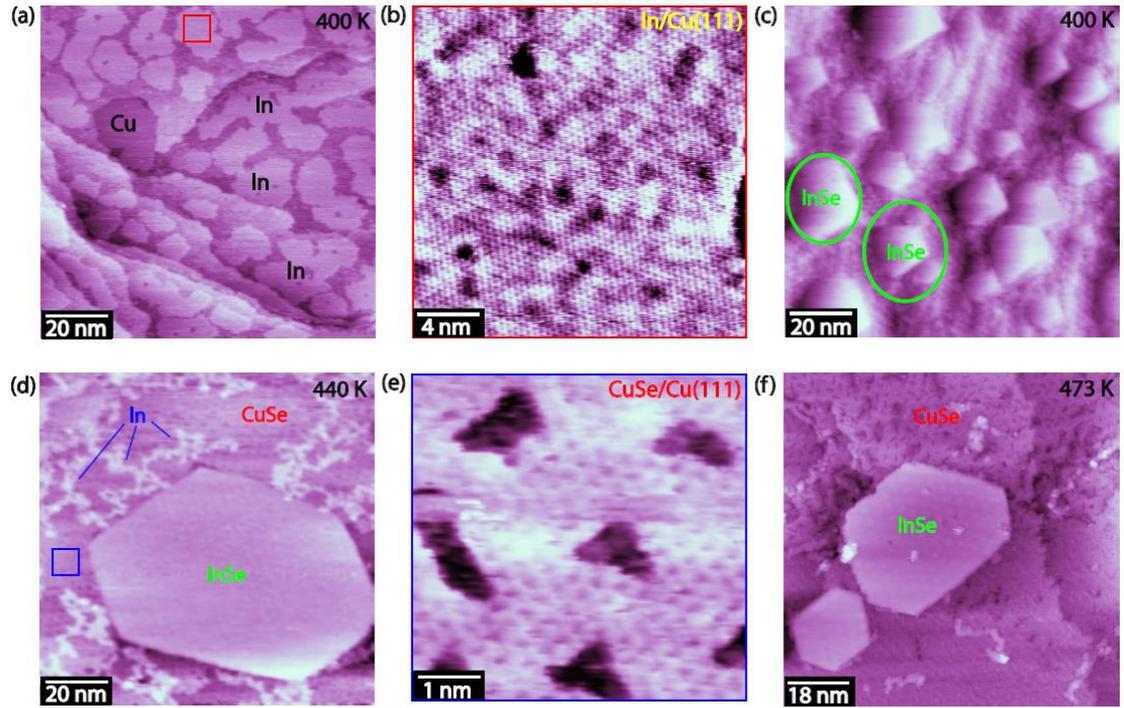

**FIG. 1** (a) Large-scale STM image of sub-monolayer In on Cu(111) at room temperature. ($V_{bias}$ = 2 V, I = 50 pA) (b) Zoom-in STM topographic image of red square marked in (a) ($V_{bias}$ = 1 V, I = 50 pA). (c) Large-scale STM image of the surface after Se deposition and annealing at 400K. ($V_{bias}$ = 1V, I = 100 pA) (d) Large-scale STM image of the surface after Se deposition and annealing at 440K. ($V_{bias}$ = 1 V, I = 100 pA). (e) Atomic STM image of the CuSe obtained from marked region in (d) ($V_{bias}$ = 1 V, I = 20 pA). (f) STM image of InSe/CuSe vertical heterostructure after annealing at 473 K ($V_{bias}$ = 2 V, I = 70 pA).

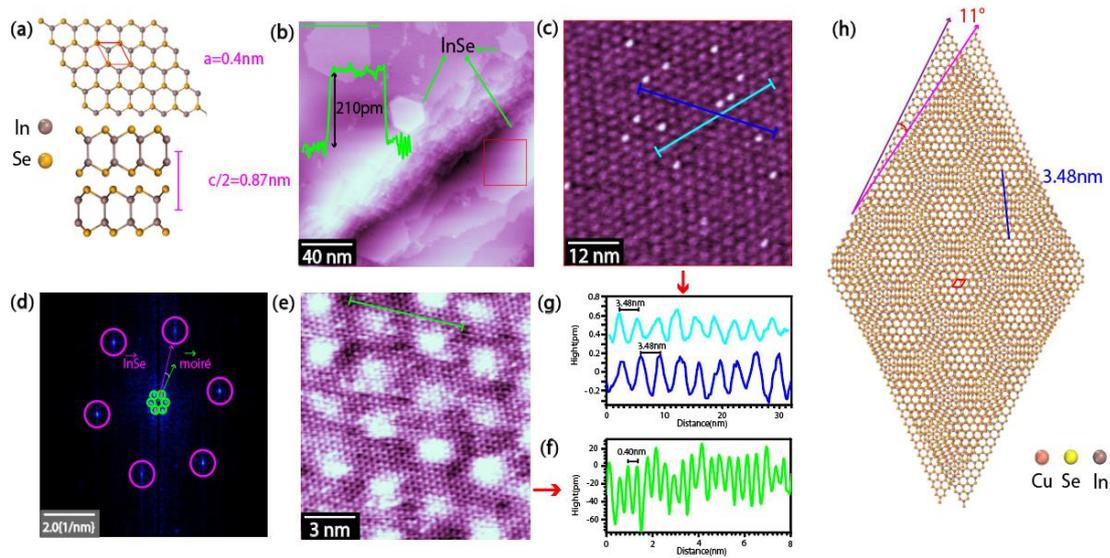

**FIG. 2** (a) Crystal structures in the top- and side-view for a InSe film, respectively. The red rhombus shows the unit cell. (b) STM morphology of InSe/CuSe vertical heterostructure. A height profile along the green line reveals the island height difference of 210 pm above the CuSe substrate ($V_{bias}$ = 1 V, I = 50 pA). (c) The zoom-in STM image of InSe island, showing a moiré superlattice. ($V_{bias}$ = 1 V, I = 110 pA) (d) 2D fast Fourier transform (FFT) pattern made from (c), revealing that reciprocal space image and detailed information for moiré periodicity and interlayer twisting. (e) High-resolution STM images of monolayer InSe film with a moire pattern with CuSe. ($V_{bias}$ = 1 V, I = 130 pA) (f) Line profile of monolayer InSe film [along the line in (e)] stands for the unit cell of monolayer InSe film, revealing the periodicity of monolayer InSe is 0.401 ± 0.002 nm. (g) Line profile of monolayer InSe film [along the lines marked in(c)], revealing the periodicity of wavelength of InSe moiré is 3.48 ± 0.05nm. (h) Schematic representation of the simulated moiré pattern of monolayer InSe and hex-CuSe.

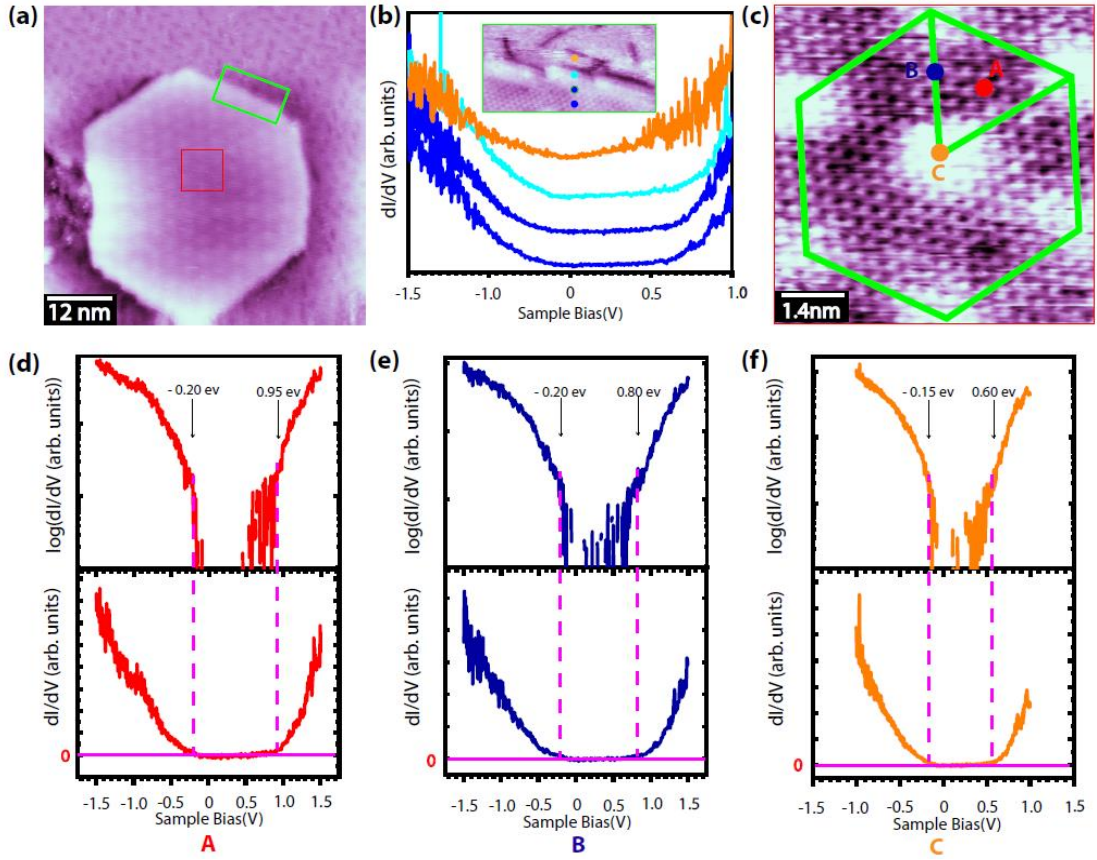

FIG. 3 (a) STM image of InSe/CuSe vertical heterostructure. (b) Serial *dI/dV* spectra collected obtained from the step edge of InSe and the substrate ($V_{bias}$ = -1.2 V, I = 90 pA) (c) High-resolution STM images of monolayer InSe films. (d)-(f) Upper panels are the logarithms of *dI/dV* spectras obtained from points A, B and C in (c), respectively, whereas the bottom panels are corresponding original *dI/dV* spectra.

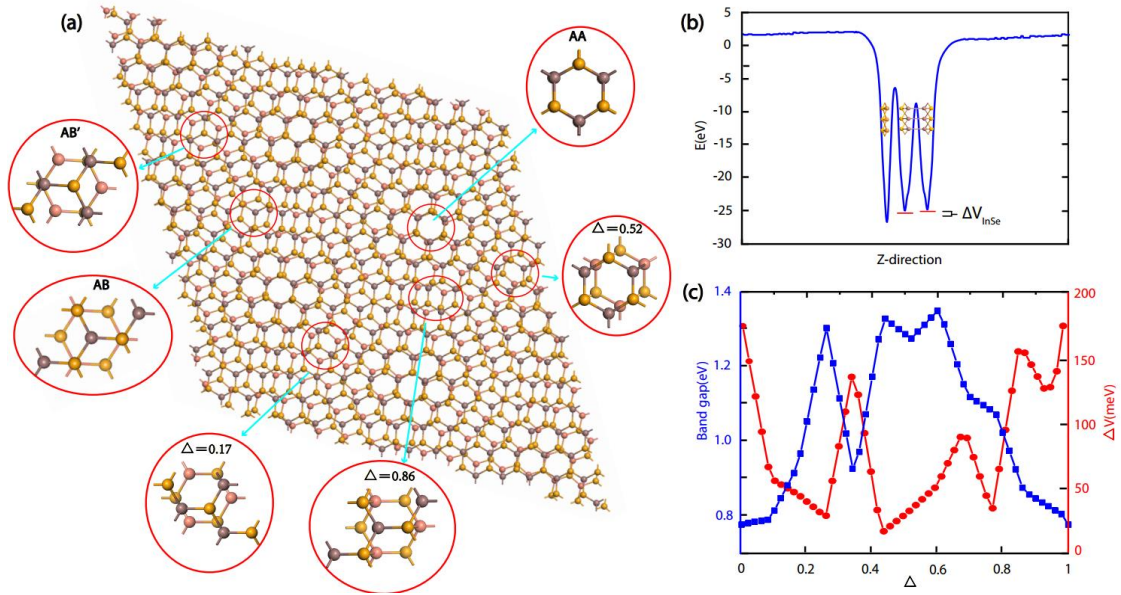

FIG. 4 (a) Moiré pattern of CuSe/InSe heterobilayer. Some local moiré stackings are magnified for more clarity. The size of moiré is equal to 6.27 nm × 6.27 nm. Some of considered stacks. Δ denotes the value of sliding monolayer InSe on CuSe in terms of lattice constant. Δ for AB, AA, and AB' stacks are 0, 0.35, and 0.7, respectively. (b) Electrostatic potential of InSe/CuSe heterobilayer along z-direction for AB stacking. The potential difference between the InSe layers is denoted by $\Delta V_{InSe}$. (c) Variation of bandgap of monolayer with lateral shift shown by blue squares. In addition, potential difference between two layers of InSe is also shown by red spots.

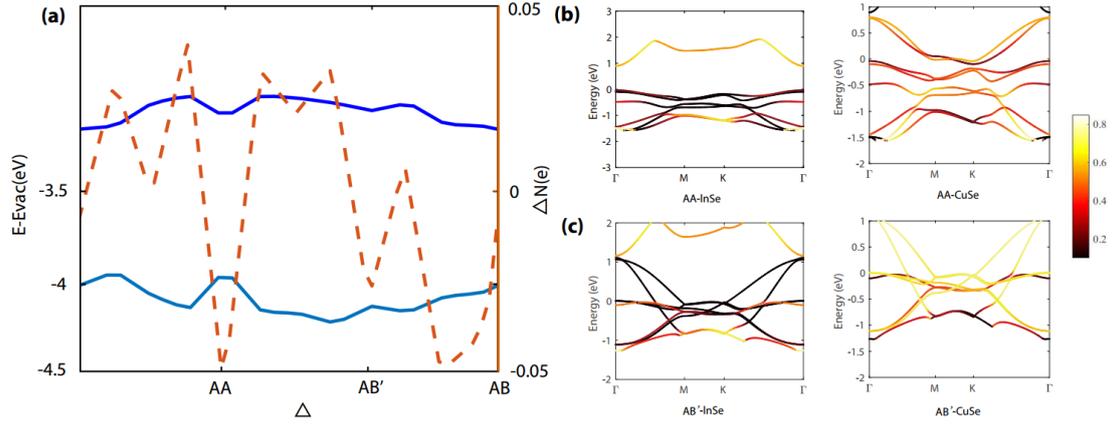

FIG. 5. (a) Charge transfer (dashed line) and the spatial variation of valence and conduction bands edge in the Γ point (solid line). (b) projected band structure of InSe (left panel) and of CuSe (right panel) for AA stacking. (c) projected band structure of InSe (left panel) and of CuSe (right panel) for AB' stacking.